\documentclass[sigconf]{acmart}
\AtBeginDocument{%
  }

\copyrightyear{2026}
\acmYear{2026}
\setcopyright{cc}
\setcctype{by}
\acmConference[WACE '26]{WebAssembly, Cloud and Edge: Shaping the Future of Computing}{December 14--18, 2026}{Tarragona, Spain}
\acmBooktitle{WebAssembly, Cloud and Edge: Shaping the Future of Computing (WACE '26), December 14--18, 2026, Tarragona, Spain}
\acmDOI{10.1145/3861044.3862585}
\acmISBN{979-8-4007-3109-9/2026/12}

\begin{document}


\title{WasmLens: Automatic Glass-Box Instrumentation for Composed WebAssembly Applications}


\author{Michiel Van Kenhove}
\orcid{0000-0003-3623-2754}
\correspondingauthor
\affiliation{%
    \institution{IDLab, Ghent University - imec}
    \city{Ghent}
    \country{Belgium}
}
\email{michiel.vankenhove@ugent.be}

\author{Ewout Verlinde}
\orcid{0009-0006-5328-0597}
\author{José Santos}
\orcid{0000-0002-6276-2057}
\affiliation{%
    \institution{IDLab, Ghent University - imec}
    \city{Ghent}
    \country{Belgium}
}

\author{Filip De Turck}
\orcid{0000-0003-4824-1199}
\author{Bruno Volckaert}
\orcid{0000-0003-0575-5894}
\author{Merlijn Sebrechts}
\orcid{0000-0002-4093-7338}
\affiliation{%
    \institution{IDLab, Ghent University - imec}
    \city{Ghent}
    \country{Belgium}
}

\renewcommand{\shortauthors}{Van Kenhove et al.}

\begin{abstract}
The WebAssembly Component Model enables language-independent composition of independently developed components, but complicates observability. Existing tracing support flattens execution hierarchy by attributing every span created inside a composition to a single host-level parent context, so execution across component boundaries cannot be reconstructed. The problem is amplified because components are typically deployed as black boxes whose internals cannot be modified or inspected by the system integrator. To address these limitations, we propose a revised \textit{wasi:otel/tracing} design in which the runtime maintains a per-component span stack, restoring parent-child relationships across component boundaries without changing component interfaces or source code. This change has since been adopted upstream. Building on this design, we present WasmLens, an automatic instrumentation tool that derives instrumentation points from a composition graph and injects tracing middleware into already composed binaries, yielding end-to-end traces for black-box compositions with no developer configuration. This automatic instrumentation raises mean request latency from $238$\,\textmu{}s to $454$\,\textmu{}s across six component boundaries. Raising the expected number of boundary crossings from 6 to 11 further increases latency to $514$\,\textmu{}s, an incremental cost of $12$\,\textmu{}s per crossing, of which $10.4$\,\textmu{}s is attributable to instrumentation. These results show that instrumentation overhead is additive in composition depth rather than proportional to baseline execution time.
\end{abstract}

\ccsdesc[500]{Software and its engineering~Runtime environments}
\ccsdesc[500]{Computer systems organization~Cloud computing}
\ccsdesc[300]{Software and its engineering~Software performance}
\ccsdesc[100]{Software and its engineering~Software testing and debugging}

\keywords{WebAssembly, Component Model, distributed tracing, OpenTelemetry, observability, auto-instrumentation, binary rewriting}

\maketitle

\section{Introduction}

Modern cloud-native applications increasingly rely on observable telemetry to reason about behavior across many independently developed and deployed components. The WebAssembly (Wasm) Component Model extends Wasm beyond the browser, enabling portable, sandboxed components to be composed through standardized interfaces~\cite{IntroductionWebAssemblyComponent}. This structured composition shares key principles with microservice architectures, particularly explicit interfaces, component autonomy, and independent evolution~\cite{jamshidiMicroservicesJourneyFar2018}, while providing lightweight isolation, portability, and compact deployment. Distributed tracing through OpenTelemetry (OTel)~\cite{OpenTelemetry} is the established means of reconstructing causal execution paths in such distributed architectures.

The Component Model breaks assumptions that distributed tracing relies on. Execution is not tied to long-running service instances: a composition may be assembled statically at build time and executed entirely within a single runtime, so architecturally significant boundaries leave no trace at the process, socket, or system-call level, placing them out of reach of zero-code approaches that observe at the kernel~\cite{OpenTelemetryEBPFInstrumentation,ashokTraceWeaverDistributedRequest2024a} or language-runtime~\cite{ZerocodeInstrumentation} boundary. Many components are also distributed as binary-only artifacts, ruling out code-level instrumentation by the integrator. Finally, the propagation mechanism currently standardized for Wasm captures only the host context present at invocation time, flattening nested spans into siblings of a single root, with more structure lost the deeper the composition.

This paper makes three contributions. We show that \textit{wasi:otel/tra-cing} cannot preserve trace hierarchy in statically composed applications, and that the obvious workaround of passing span context through WIT interfaces is unacceptable in a black-box setting (Section~\ref{sec:motivation}). We propose a revised interface in which the runtime maintains a per-component span stack, replacing the static \textit{outer-span-context} accessor with a dynamic \textit{current-span-context} query. We implement it in wasmCloud's~\cite{wasmCloudWebAssemblyApplicationPlatform} runtime, and our proposed interface change has since been adopted upstream~\cite{ReplaceOuterspancontextCurrentspancontext} in the \textit{wasi:otel} specification (Section~\ref{sec:design}). We then present WasmLens, an open-source tool~\cite{IdlabdiscoverWasmLens2026}, which derives instrumentation points from a composition graph and injects OTel tracing into an already composed binary without source code access or manual configuration (Section~\ref{sec:wasmlens}). We then show that its overhead grows linearly with the number of boundaries a request crosses (Section~\ref{sec:eval}). Finally, Section~\ref{sec:related-work} positions our approach against existing zero-code instrumentation, and Section~\ref{sec:conclusion} concludes the paper.

\section{Background and Motivation}\label{sec:motivation}

\subsection{Composing WebAssembly Components}\label{sec:demo}

The Component Model raises the abstraction level at which Wasm binaries interact, using the WebAssembly Interface Types (WIT) language to define safe, language-agnostic interfaces. Components can be linked statically at build time with tools such as wac~\cite{BytecodeallianceWac2026}, producing a single composed component. This linking process automatically generates lightweight intermediary shim components that reconcile interface mismatches.

\begin{figure}
    \centering
    \includegraphics[width=\columnwidth]{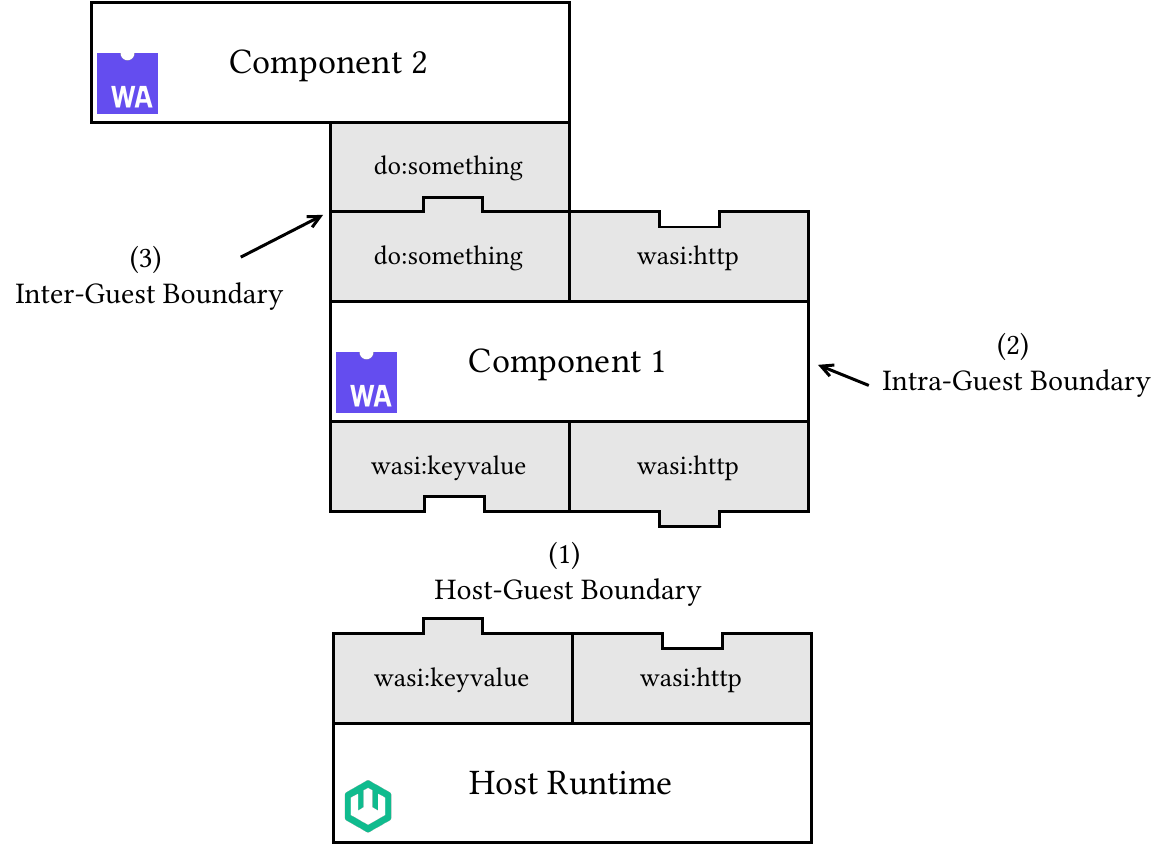}
    \caption{The host-guest (1), intra-guest (2), and inter-guest (3) context boundaries in a composed WebAssembly application.}\label{fig:wasm-boundaries}
    \Description[A block diagram of a WebAssembly application in which a host runtime, one component stacked on it, and a second composed component are separated by three numbered context boundaries.]{A block diagram built from rectangles joined by interlocking puzzle-piece connectors, each connector labeled with a WIT interface name. At the bottom, a box labeled "Host Runtime" exposes two interfaces along its top edge: wasi:keyvalue and wasi:http. Directly above it sits a box labeled "Component 1", marked with the WebAssembly logo, whose lower edge carries matching wasi:keyvalue and wasi:http connectors that interlock with the host's. The seam between the two is annotated "(1) Host-Guest Boundary". Component 1's upper edge carries two further interface blocks, labeled do:something and wasi:http. Above and to the left, a second box labeled "Component 2", also marked with the WebAssembly logo, exposes a do:something interface on its lower edge, which interlocks with the do:something connector of Component 1. An arrow from the left labeled "(3) Inter-Guest Boundary" points at this seam. A third arrow, from the right, labeled "(2) Intra-Guest Boundary", points at the body of Component 1 itself, indicating transitions between execution contexts within a single component rather than at any interface seam.}
\end{figure}

Context propagation must account for three execution boundaries, shown in Figure~\ref{fig:wasm-boundaries}: the \textbf{host-guest} boundary (1) between a guest and host-provided capabilities such as WASI interfaces; the \textbf{intra-guest} boundary (2) between internal transitions within a single component; and the \textbf{inter-guest} boundary (3) between composed components. Inter-guest boundaries are either static, resolved at composition time via shims, or dynamic, mediated by the runtime. Static inter-guest boundaries are the hard case because at runtime they are indistinguishable from ordinary intra-guest execution.

\begin{figure}
    \centering
    \includegraphics[width=\columnwidth]{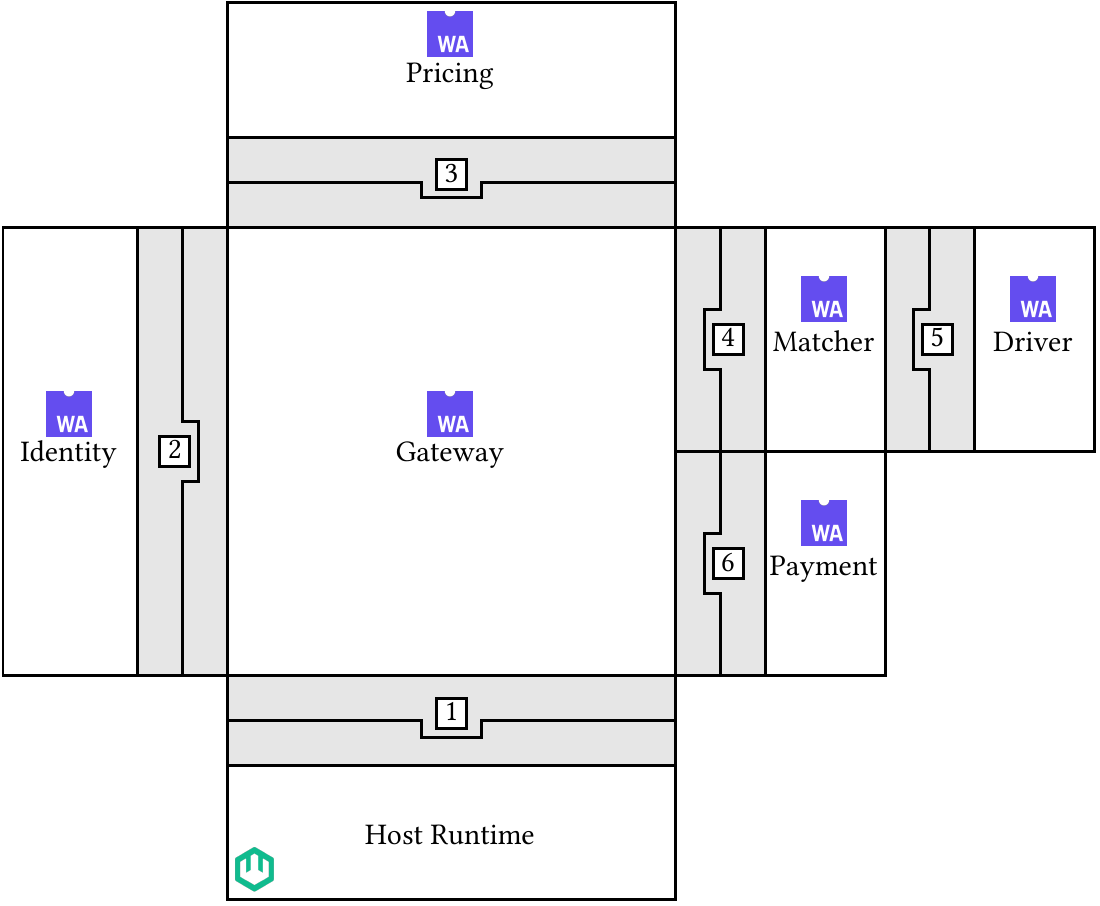}
    \caption{The statically composed ride-hailing application, with six components linked by wac into a single guest component. A request crosses the host-guest boundary (1), then the inter-guest boundaries to identity (2), pricing (3), matcher (4), driver (5), and payment (6).}\label{fig:demo}
    \Description[Block diagram of the six-component ride-hailing composition, with numbered bands marking the boundaries a request crosses from the host runtime through the gateway to the identity, pricing, matcher, driver, and payment components.]{A block diagram of seven labeled boxes, each of the six WebAssembly components marked with a Wasm logo. The gateway is at the center. The host runtime is below it, identity to its left, pricing above it, and matcher and payment to its right, with the driver component placed beyond the matcher on the far right. Each pair of adjacent boxes is separated by a shaded band representing the interface boundary between them, and each band carries a numbered marker giving the order in which a request crosses it. Marker 1 lies between the host runtime and the gateway. Marker 2 lies between the gateway and identity, marker 3 between the gateway and pricing, marker 4 between the gateway and matcher, marker 5 between the matcher and driver, and marker 6 between the gateway and payment. Boundary 1 is a host-guest boundary; boundaries 2 to 6 are static inter-guest boundaries created by composition.}
\end{figure}

A composed ride-hailing application, shown in Figure~\ref{fig:demo}, serves as a running example throughout the paper and as the evaluated workload in Section~\ref{sec:eval}. Six components are statically composed with wac: a \emph{gateway} exposing \textit{wasi:http/incoming-handler} and calling, in turn, an \emph{identity} component that validates a bearer token, a \emph{pricing} component that computes a fare, a \emph{matcher} that selects candidate drivers, and a \emph{payment} component that returns a payment URL. The matcher invokes a \emph{driver} component in a retry loop, repeating until a candidate accepts. Every other interaction occurs exactly once per request. Composition is nested, with driver linked into matcher first and the result composed with the remaining components. Only \textit{wasi:http/incoming-handler} is exported at the top level, so the runtime sees the whole application as one opaque guest instance. We deploy locally with wash~\cite{WashWasmShell2026}, which embeds a Wasmtime~\cite{BytecodeallianceWasmtime2026} based host runtime.

\subsection{Tracing Across Composed Boundaries}\label{sec:challenge-hierarchy}

OTel represents execution as traces of spans connected by parent-child relationships, maintained through context propagation~\cite{ContextPropagation}. The \textit{wasi:otel/tracing} interface~\cite{WebAssemblyWasiotel2026,UnlockingOtelWebAssembly2025} standardizes emission of OTel traces from Wasm guests. It provides functions for creating and completing spans, and, in version 0.2.0-rc.1, for accessing the active parent host span context via \textit{outer-span-context} at component entry. This design assumes that a host-provided span context is available at invocation time and can be used as the parent context for subsequently created spans.

In a static composition the runtime treats the entire composite as one guest instance, so \textit{outer-span-context} returns the span context active on the host side at that moment. That context is neither derived from nor affected by spans emitted inside the composition during execution. Using it as the parent of every new span produces a flat hierarchy in which all component spans attach directly to the host's entry span: when the gateway calls the matcher, the matcher's span does not nest beneath the gateway's, and both become siblings of the root. The structure needed to reconstruct execution order is absent, and causal reasoning across components becomes unreliable. The same gap appears without any boundary crossing, since the interface exposes no runtime-managed notion of an active span: nested spans created inside a single component cannot resolve their parent either.

The direct workaround is to propagate context explicitly through WIT, extending every function signature with a \textit{span-context} parameter. However, this is unattractive for three reasons. It violates separation of concerns, since business-logic interfaces become dependent on the telemetry stack. It also propagates, in the sense that every caller along every path must pass a context object through. And it forces recompilation of components with no observability requirement of their own, which, in a composition assembled from third-party binaries, is exactly the set the integrator cannot rebuild.

\subsection{Observing Black-Box Components}\label{sec:challenge-blackbox}

Components are often independently developed, versioned, and distributed as binary artifacts, integrated purely through interface definitions. The Component Model deliberately reinforces this, since composition is established by interface compatibility rather than access to implementation details. Without explicit instrumentation, execution inside a component is therefore not observable from outside. Such components become dark nodes in an otherwise well-structured composition graph, and trace continuity breaks at every non-participating component.

Instrumenting every component manually is not a general solution. Integrators often lack source access to the components they compose. Embedding tracing logic entangles operational infrastructure with business functionality, expanding the WIT interfaces a component imports or exports with propagation primitives. Instrumentation effort also competes with feature work, so coverage remains incomplete even where standardized APIs exist~\cite{ashokTraceWeaverDistributedRequest2024a,santanaTransparentTracingMicroservicebased2019c}.

Together with the trace flattening issue above, these constraints motivate a shift from manually instrumented black-box compositions to automatically instrumented \emph{glass-box} compositions, in which tracing is introduced after composition but before deployment.

\section{Ambient Context Propagation}\label{sec:design}

Restoring trace hierarchy without polluting interfaces requires three things. Each component must resolve its immediate caller as parent, rather than the host root span. No context object may appear in a business-logic WIT signature. And a guest must be able to query the active span at any point during execution, as native OTel SDKs allow.

We therefore replace the static \textit{outer-span-context} accessor with a dynamic \textit{current-span-context} query backed by host-managed ambient state. The host maintains a logical span stack per component instance: it pushes the incoming context on \textit{on-start}, returns the top of the stack on \textit{current-span-context}, and removes the corresponding entry on \textit{on-end}. If a gateway component starts a span and then calls a service component, the callee's \textit{current-span-context} then returns the gateway's active span. When the stack is empty, the host falls back to its own ambient context, reproducing the previous behavior and preserving continuity with the incoming request. Following design discussion~\cite{LimitationWasiOtel} with the \textit{wasi:otel} maintainers, the static accessor was found redundant once ambient state exists, and the replacement has been adopted upstream~\cite{ReplaceOuterspancontextCurrentspancontext}.

We implement the design in the host runtime that wasmCloud's wash embeds. Host-side implementations of WIT interfaces are shared across all components consuming them, so per-component state must be isolated explicitly. We therefore extend the runtime's \textit{ComponentContext} with a span stack keyed per component invocation rather than per component, which avoids shared mutable state between concurrent invocations. Stack discipline is sound because component invocations execute synchronously, so \textit{on-start} and \textit{on-end} events are strictly ordered within one execution path. Under the optional threading proposal the stack would need to be initialized per thread. For dynamically linked components, whose boundaries the runtime mediates across process or network boundaries, the host instead encodes the active context into transport metadata before dispatch and reinitializes the local span stack on receipt, a form of propagation wasmCloud and Spin already implement~\cite{ObservabilityOpenTelemetryWasmCloud2026,ObservingApplications2023}.

Running the example of Section~\ref{sec:demo} on the modified runtime yields the expected hierarchy. The gateway span becomes the parent of the identity, pricing, matcher, and payment spans, and matcher the parent of driver, instead of the six sibling spans the original interface produces. This result is obtained without modifying any WIT interface or application call signature.

\section{WasmLens}\label{sec:wasmlens}

Correct propagation makes hierarchical traces possible, but it does not make them automatic. WasmLens~\cite{IdlabdiscoverWasmLens2026} provides this automation by injecting tracing at the component boundaries of an already composed binary, using Splicer~\cite{gilbertWriteOnceInstrument2026,evilgEjrgilbertSplicer2026} as its transformation layer.

\subsection{Composition Rewriting}

Splicer rewrites Wasm component graphs through declarative injection rules. Given a target composition, a middleware binary, and a rule set, it decomposes the component graph, inserts adapter components at selected interface edges, and recomposes a valid Wasm binary. WasmLens targets Splicer's Tier 1 contract, which exposes minimal invocation metadata, notably a \textit{call-id}, without access to function arguments or payloads. A \textit{call-id} is enough to correlate the entry and exit of the same invocation. The adapter preserves the original WIT signature while routing execution through Tier 1 hooks, so the same tracing middleware remains reusable across arbitrary interfaces. The cost is indirection. An uninstrumented boundary is represented by a single generated shim, whereas an instrumented one becomes three shimmed edges: component to adapter, adapter to middleware, and adapter to downstream component.

\subsection{Automatic Boundary Detection}

Splicer alone requires developers to enumerate the boundaries to instrument. That is reasonable for targeted middleware but defeats the purpose of auto-instrumentation: it presumes detailed knowledge of the composition graph, including the internal structure of third-party components that should remain opaque, and must be maintained as the composition evolves. WasmLens instead derives candidate boundaries from the composition structure itself, using the graph representation produced by cviz~\cite{CosmoniclabsCviz2026}. We classify instrumentation points into two categories: top-level exports of the root component, which represent system entry points such as HTTP handlers; and internal import relationships between components or between a component and the host. Both correspond to control-flow transitions and are therefore natural span boundaries. Traversing exported interfaces and component imports produces a complete candidate set, which WasmLens emits as Splicer injection rules. By default every such edge is instrumented, so coverage extends through nested compositions without any hand-written rule.

\subsection{Tracing Middleware}

The middleware constructs OTel spans around each intercepted invocation through two Tier 1 hooks. On entry the middleware retrieves the parent span via \textit{current-span-context} and builds a new child span that inherits its trace identifier and trace state, recording a start timestamp and a name derived from the interface and function being called. Because the Tier 1 hooks are stateless, this span is held in an in-flight map keyed by \textit{call-id}. On return, the matching entry is retrieved and a complete \textit{SpanData} is exported via \textit{on-end}.

Applying WasmLens to the example binary of Section~\ref{sec:demo} yields a trace structurally equivalent to the manually instrumented variant: each detected boundary appears as a span named by its fully qualified interface and function identifier, hierarchy is preserved through \textit{current-span-context}, and host and guest execution remain distinguishable through reported service names. WasmLens therefore achieves semantic parity with manual instrumentation while eliminating boundary configuration entirely.

\section{Evaluation Results}\label{sec:eval}

All experiments run on a Lenovo LOQ 15IRX10 with an Intel Core i7-13700HX and 24\,GB RAM running Arch Linux, using wasmCloud \textit{wash-runtime} v2.1.0 extended with the interface of Section~\ref{sec:design} and a local OTel collector. Load is generated with wrk~\cite{glozerWgWrk2026}, ten seconds per run, with the load generator pinned to a disjoint CPU range so that the measured overhead reflects instrumentation rather than client-server contention. We compare three variants of the application of Section~\ref{sec:demo}: uninstrumented, manually instrumented, and automatically instrumented with WasmLens.

\subsection{Per-Boundary Overhead}\label{sec:eval:hops}

With the six boundary crossings of Figure~\ref{fig:demo}, the uninstrumented baseline has a mean latency of $238$\,\textmu{}s, manual instrumentation $300$\,\textmu{}s, and WasmLens $454$\,\textmu{}s. Dividing the added latency by the six boundaries gives a first estimate of the per-boundary cost: $\approx 10$\,\textmu{}s for manual instrumentation and $\approx 36$\,\textmu{}s for automatic instrumentation. Expressed as a ratio, automatic instrumentation adds 91\% to the baseline execution time. That framing is potentially misleading though: the uninstrumented baseline is intentionally lightweight, so any fixed overhead looks large relative to it.

To test whether the overhead is indeed fixed per boundary, we vary the number of boundary crossings without changing the actual work performed. The matcher component queries one driver at a time and retries on rejection. The probability that a driver accepts is $p=1/n$. Increasing $n$ therefore increases matcher-to-driver calls and thus adds inter-component boundary crossings without altering the remaining execution path. Let $X\sim\mathrm{Geom}(p)$ denote the number of driver attempts, then $\mathbb{E}[X]=1/p=n$. With five fixed boundary crossings outside the retry loop, the expected number of boundary crossings per request is $\mathbb{E}[B]=5+n$.

Table~\ref{tab:hop_scaling} reports the results. Latency grows linearly with $n$ in all three variants. Each additional matcher-to-driver crossing costs $1.6$\,\textmu{}s uninstrumented, $6.7$\,\textmu{}s with manual instrumentation, and $12.0$\,\textmu{}s with WasmLens ($R^2 = 0.77$, $0.93$, and $0.99$ respectively).

Even without tracing, a crossing is not free. The $1.6$\,\textmu{}s slope is the cost of the shims the Component Model inserts between components. Subtracting it leaves the cost of tracing alone: about $5.1$\,\textmu{}s per boundary for manual instrumentation and $10.4$\,\textmu{}s for WasmLens. Both variants make the same three \textit{wasi:otel} host calls per boundary. The extra cost aligns with Splicer's adapter-based transformation, which adds indirection layers at every instrumented boundary (Section~\ref{sec:wasmlens}).

These numbers are lower than the $10$ and $36$\,\textmu{}s estimated from the baseline, because not every boundary costs the same. A boundary's cost depends on how much metadata its host calls carry, and for WasmLens on how many shims it expands into. The matcher-to-driver call carries a small payload, so it sits at the cheap end. What both measurements agree on is that instrumentation adds a cost per boundary crossed, and that this cost does not grow as compositions get deeper.

\begin{table}[t]
  \caption{Latency and throughput at concurrency $c=1$ against composition depth, where $\mathbb{E}[B] = 5 + n$ is the expected number of boundary crossings per request. Latency is the mean per-request latency over a ten-second run. Throughput is given in requests per second (RPS). Three of the six measured values of $n$ are shown.}\label{tab:hop_scaling}
  \begin{tabular}{lrrrr}
    \toprule
    Variant & $n$ & $\mathbb{E}[B]$ & Latency (\textmu{}s) & RPS\\
    \midrule
    Uninstrumented  & 1 & 6  & 238 & 4146\\
                    & 3 & 8  & 242 & 4059\\
                    & 6 & 11 & 245 & 4030\\
    \addlinespace
    Manual          & 1 & 6  & 300 & 3272\\
                    & 3 & 8  & 308 & 3192\\
                    & 6 & 11 & 331 & 2941\\
    \addlinespace
    Auto (WasmLens) & 1 & 6  & 454 & 2141\\
                    & 3 & 8  & 479 & 2031\\
                    & 6 & 11 & 514 & 1896\\
    \bottomrule
  \end{tabular}
\end{table}

\subsection{Overhead Under Load}

The measurements up until now send one request at a time. We now check whether instrumentation costs proportionally more as concurrent load increases. To do so, we vary the client-side concurrency $c \in \{1,2,4,8,16,32\}$. Throughout these runs, we keep $p$ fixed at 1, so that the first driver always accepts and no matcher-to-driver retries occur. Every request therefore follows the same deterministic path with exactly $B = 6$ boundary crossings.

\begin{figure}
    \centering
    \includegraphics[width=\columnwidth]{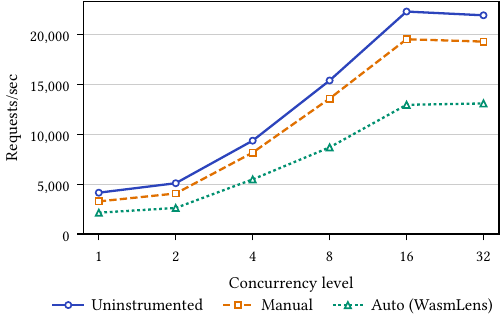}
    \caption{Throughput as client concurrency $c$ increases, with $p=1$ so that every request makes $B=6$ boundary crossings.}\label{fig:concurrency}
    \Description[Line chart: throughput of three variants rises with concurrency and plateaus at 16, with Auto lowest throughout]{Line chart of throughput in requests per second (y-axis, 0 to about 22,000) against client concurrency level (x-axis: 1, 2, 4, 8, 16 and 32, evenly spaced on a logarithmic scale). There are three series: Uninstrumented (solid line, circle markers), Manual (dashed line, square markers) and Auto (WasmLens) (dotted line, triangle markers). All three rise gently from concurrency 1 to 2, climb more steeply from 2 to 16, and flatten between 16 and 32. Uninstrumented is highest at every level. It goes from about 4,200 requests per second at concurrency 1 to 22,250 at 16 and 21,900 at 32. Manual sits slightly below it, going from about 3,300 to 19,500 and 19,250. Auto is lowest, going from about 2,100 to 12,950 and 13,050. The lines never cross. The absolute gap between Auto and the other two variants widens as concurrency increases.}
\end{figure}

Figure~\ref{fig:concurrency} shows that the uninstrumented baseline maintains the highest throughput, with the manually and automatically instrumented variants progressively lower, reflecting their additional per-request overhead compared to baseline. Despite these absolute differences, all three behave the same way as load increases. Throughput grows with concurrency up to $c \approx 16$, beyond which additional concurrency no longer increases throughput and the latency rises instead. At $c=32$ the three variants sustain 21,909, 19,276, and 13,081 requests per second.

Instrumentation therefore costs a roughly constant fraction of throughput at every concurrency level rather than a growing one. WasmLens sustains between 51\% and 60\% of the baseline throughput across the range, and manual instrumentation between 79\% and 88\%, with no downward trend as load increases. All three variants also saturate at the same concurrency level, so instrumentation is not what limits throughput under load.

\subsection{Binary Size}\label{sec:eval:bloat}

Instrumentation has little effect on the binary size. The composed binary grows from 2372\,KiB uninstrumented to 2420\,KiB with manual instrumentation ($+2.0\%$) and 2400\,KiB with WasmLens ($+1.2\%$). The difference between the two approaches is negligible, though it is consistent with WasmLens including the tracing logic once as shared middleware rather than linking it into each component separately.

\section{Related Work}\label{sec:related-work}

Zero-code instrumentation removes the need for developers to insert spans and propagate context by hand. Two general-purpose families dominate, differing in their \emph{observation boundary}, which determines what survives in a WebAssembly composition. We then discuss the only Wasm-specific approach we are aware of.

\paragraph{Language-level instrumentation}
OTel provides automatic instrumentation for multiple programming languages and frameworks, such as Go, Java, Python, JavaScript, PHP, and .NET by hooking method invocations, framework lifecycle events, and library calls within the application's language runtime~\cite{ZerocodeInstrumentation}. This approach yields rich semantic detail, but the enabling mechanisms, including monkey patching, bytecode rewriting, and runtime agents, depend on features of specific language runtimes and are unavailable in Wasm. Compiling existing OTel SDKs to Wasm is itself obstructed by system-interface assumptions and platform-specific hooks~\cite{UnlockingOtelWebAssembly2025}.

\paragraph{Kernel-level instrumentation}
eBPF provides an event-driven mechanism for attaching programs to kernel hooks, enabling system tracing and network observability without modifying the monitored applications~\cite{craunEliminatingEBPFTracing2024,rezvaniCharacterizingInKernelObservability2024}. OTel eBPF Instrumentation~\cite{OpenTelemetryEBPFInstrumentation} captures protocol-level interactions such as HTTP and gRPC, while MiSeRTrace~\cite{vMiSeRTraceKernellevelRequest2022} associates kernel events with individual requests, and TraceWeaver~\cite{ashokTraceWeaverDistributedRequest2024a} reconstructs request paths from externally collected span timings and call-graph information. Building on timing-based reconstruction, CrossTrace~\cite{phanCrossTraceEfficientCrossThread2025} correlates timestamped in-gress and egress spans within a service without relying on OS-thread affinity, an assumption that can fail when requests migrate between threads or when a thread handles multiple logical requests concurrently~\cite{shenNetworkCentricDistributedTracing2023, ashokTraceWeaverDistributedRequest2024a}. Across services, CrossTrace propagates span identifiers through TCP header options.

None of this transfers to composed Wasm. A Wasm runtime mediates all system interaction, so many component-level operations are multiplexed onto one runtime thread and surface to the kernel as a single syscall sequence. Statically composed boundaries are resolved at build time and crossed by in-process calls that generate no socket or syscall activity at all, removing the very signal these systems correlate on. A single runtime instance may also execute unrelated components concurrently, interleaving their events into one kernel-observable stream with no identifier to attribute them.

\paragraph{Wasm-specific tracing}
WasmTracer~\cite{nakamuraWasmTracerImproving2025} moves the observation boundary closer to the guest by observing WASI-mediated system calls through eBPF and constructing OTel spans in a user-space agent. Because spans are derived from monitored syscall activity, execution that does not result in such interactions is not directly captured by the approach. Unlike instrumentation inside the guest, the observation boundary remains outside the Wasm module. Although the paper outlines an implementation based on eBPF tracing and OTel integration, it reports that implementation and evaluation were still ongoing at the time of publication.

Existing systems optimize either for low-level system breadth or for application-level semantic depth, but not both in a Wasm setting. WasmLens shifts the observation boundary to the composition graph itself, so inter-component interactions are observed directly rather than inferred from kernel- or WASI-level events. It takes one property from each family. Like language-level instrumentation it records the interfaces and functions being called, rather than reconstructing structure from lower-level events. Like kernel-level instrumentation it needs no cooperation from the code it observes and works regardless of the language a component was written in, since instrumentation attaches to the compiled interface boundary that the Component Model defines independently of any source language.

\section{Conclusion}\label{sec:conclusion}

Composed WebAssembly applications lose their trace hierarchy at statically composed boundaries, and the components themselves often cannot be instrumented by the integrator who composes them. We addressed the first with a revised \textit{wasi:otel/tracing} design in which the runtime maintains a per-component span stack, restoring parent and child relationships without requiring changes to any component interface. This change has been adopted upstream. We addressed the second with WasmLens, which derives instrumentation points from a composition graph and injects tracing middleware into already composed binaries, producing traces equivalent to manual instrumentation without configuration or access to their source code.

Instrumentation adds a fixed cost per boundary crossing rather than a cost proportional to the time a request takes. Whether that cost matters depends on how much work the components themselves do. For example, when a request spends milliseconds in application logic or waiting on host calls, tens of microseconds per boundary can be considered negligible. On the other hand, when a request performs little actual computation and primarily crosses boundaries, as in our benchmark, the overhead becomes visible. That overhead is offset by the ability to trace components that would otherwise remain unobservable, since WasmLens can instrument them without source code changes or manual context propagation. The two approaches can also be combined, with an integrator instrumenting the components they control by hand and employing WasmLens for the other, possibly third-party, components.

Several directions remain open as future work. Our evaluation covers one application on one machine, so measurements across more compositions would further validate the per-boundary figures we report. Additionally, WasmLens targets Splicer's Tier 1 middleware, which exposes only interface name, function name, and call identifier. Tier 2 middleware grants read-only access to function parameters, which would allow richer span attributes derived from runtime arguments, but introduces challenges related to Application Binary Interface (ABI) diversity, schema uncertainty, and the safe handling of potentially sensitive data. That extension therefore requires both robust parameter extraction and appropriate redaction strategies. A second direction is configurability. WasmLens currently instruments every detected boundary, and letting users selectively include or exclude interfaces would reduce telemetry noise in large compositions.

\begin{acks}

The authors thank Elizabeth Gilbert (Carnegie Mellon University) for valuable discussions on WebAssembly component interposition and for their work on Splicer, which directly influenced WasmLens.

This work received funding from the \grantsponsor{snsju-horizon}{Smart Networks and Services Joint Undertaking (SNS JU) under the Horizon Europe research and innovation program of the European Union}{https://doi.org/10.3030/101139067} under Grant Agreement number \grantnum[https://doi.org/10.3030/101139067]{snsju-horizon}{101139067} (\href{https://elasticproject.eu/}{https://elasticproject.eu/}). Expressed views and opinions are however those of the authors only and do not necessarily reflect those of the European Union. Neither the European Union nor the granting authority can be held responsible for them.

José Santos is funded by the \grantsponsor{jose-fwo}{Research Foundation Flanders (FWO)}{https://www.fwo.be/en/}, grant number \grantnum{jose-fwo}{1253226N}.
\end{acks}

\bibliographystyle{ACM-Reference-Format}
\bibliography{references}










\end{document}